\documentclass[preprint,11pt,authoryear]{elsarticle}
\pdfoutput=1
\usepackage{graphicx}
\usepackage{amsmath}
\usepackage{esint}
\usepackage{amssymb}
\usepackage{lineno}
\usepackage[T1]{fontenc}
\usepackage[utf8]{inputenc}
\usepackage[english]{babel}
\usepackage[]{algorithm}
\usepackage[noend]{algpseudocode}

\biboptions{square}

\usepackage[scaled]{helvet}

\usepackage{units}
\usepackage[usenames, dvipsnames]{xcolor}

\usepackage{soul}
\usepackage{placeins}

\usepackage{nameref}
\usepackage[pdftex,breaklinks=true,colorlinks=true,linkcolor=blue,citecolor=blue,urlcolor=blue,filecolor=blue,pdffitwindow,backref=true,pagebackref=false,bookmarks=true,bookmarksopen=true,bookmarksnumbered=true]{hyperref}
\usepackage[plain]{fancyref}
\usepackage{array}
\usepackage{multirow}

\Frefformat{plain}{\fancyreffiglabelprefix}{Fig~#1}
\newcommand*{\Frefsecshortname}{Section}%
\Frefformat{plain}{\fancyrefseclabelprefix}{\Frefsecshortname~#1}
\Frefformat{plain}{\fancyrefeqlabelprefix}{Eq~(#1)}

\usepackage{tabularx}
\usepackage{colortbl}
\usepackage{morefloats}

\usepackage{float}
\floatstyle{plaintop}
\restylefloat{table}

\makeatletter
\def\ps@pprintTitle{%
 \let\@oddhead\@empty
 \let\@evenhead\@empty
 \def\@oddfoot{}%
 \let\@evenfoot\@oddfoot}
\makeatother

\usepackage[exponent-product = \cdot]{siunitx}
\usepackage[colorinlistoftodos]{todonotes}
\usepackage[normalem]{ulem} 
\usepackage{caption}
\usepackage{booktabs} 
\graphicspath{{Figures/}} 
\usepackage{caption}
\usepackage{subcaption}

\begin{document}

\begin{frontmatter}


\title{Computing extracellular electric potentials from neuronal simulations} 



\author{Torbj\o{}rn V. Ness\corref{equal} \fnref{label1}}
\author{Geir Halnes\corref{equal} \fnref{label1}}
\author{Solveig N\ae{}ss\fnref{label2}}
\author{Klas H. Pettersen \fnref{label3}}
\author{Gaute T. Einevoll\corref{cor1}\fnref{label1,label4}}

\address[label1]{Faculty of Science and Technology, Norwegian University of Life Sciences, {{\AA}}s, Norway}
\address[label2]{Department of Informatics, University of Oslo, Oslo, Norway}
\address[label3]{Norwegian Artificial Intelligence Research Consortium, Oslo, Norway}
\address[label4]{Department of Physics, University of Oslo, Oslo, Norway}

\cortext[equal]{These authors have contributed equally to this work}
\cortext[cor1]{correspondence: \href{gaute.einevoll@nmbu.no}{gaute.einevoll@nmbu.no}}


\begin{abstract}
Measurements of electric potentials from neural activity have played a key role in neuroscience for almost a century, and simulations of neural activity is an important tool for understanding such measurements. Volume conductor (VC) theory is used to compute extracellular electric potentials such as extracellular spikes, MUA, LFP, ECoG and EEG surrounding neurons, and also inversely, to reconstruct neuronal current source distributions from recorded potentials through current source density methods. In this book chapter, we show how VC theory can be derived from a detailed electrodiffusive theory for ion concentration dynamics in the extracellular medium, and show what assumptions that must be introduced to get the VC theory on the simplified form that is commonly used by neuroscientists. Furthermore, we provide examples of how the theory is applied to compute spikes, LFP signals and EEG signals generated by neurons and neuronal populations.
\end{abstract}

\begin{keyword}
extracellular potentials \sep LFP \sep EEG \sep ECoG \sep electrodiffusion \sep neuronal simulation \sep MUA

\end{keyword}

\end{frontmatter}

\tableofcontents


\section{Introduction}
\label{sec:introduction}

\begin{figure}[!ht]
\begin{center}
\includegraphics[width=0.8\textwidth]{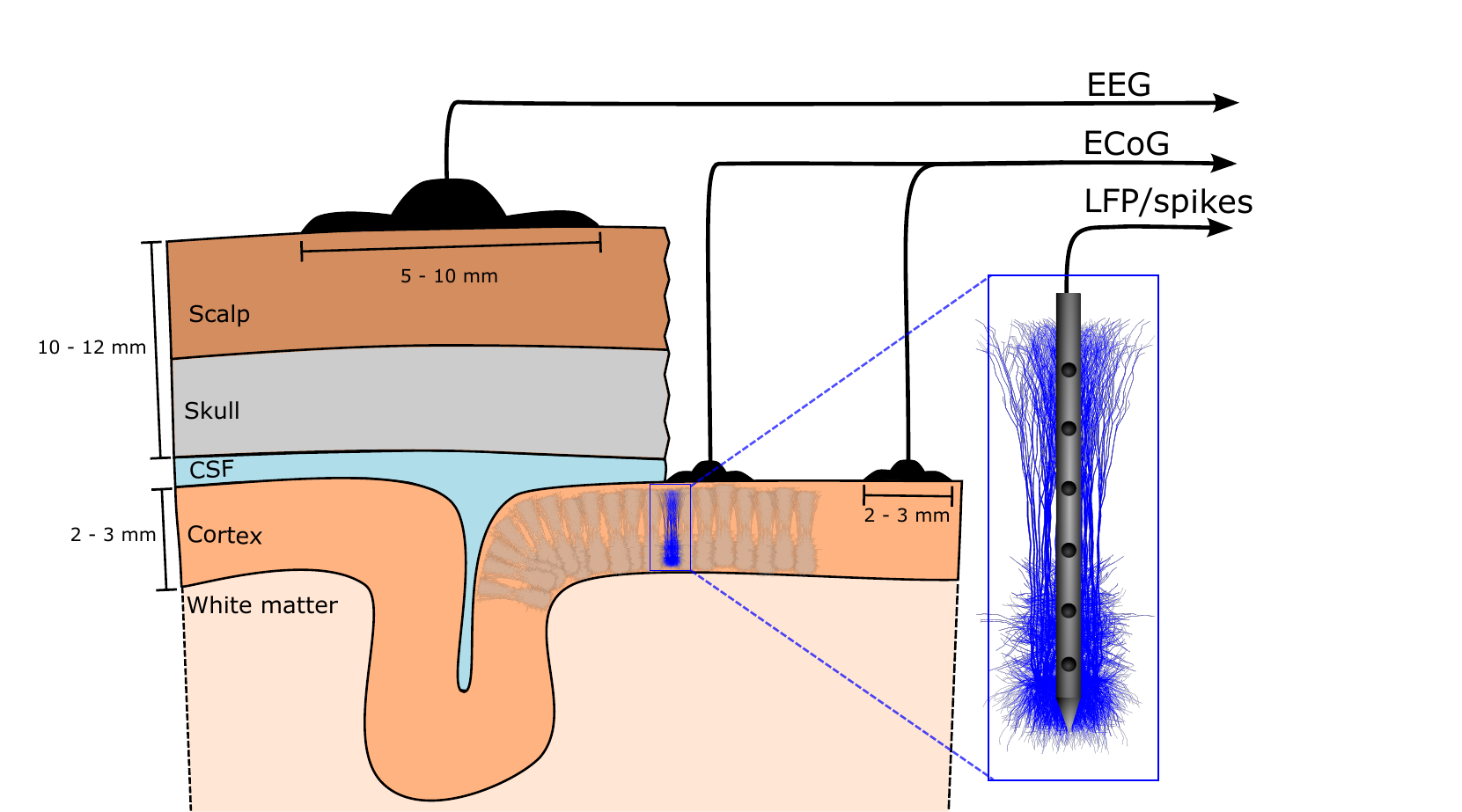}
\end{center}
\caption{\textbf{LFP, ECoG and EEG.} The same basic building blocks, that is, currents caused by large numbers of synaptic input are contributing to several different measurable signals.
}
\label{fig:multimodal}
\end{figure}

Arguably, most of what we have learned about the mechanisms by which neurons and networks operate in living brains comes from recordings of 
extracellular potentials. In such recordings, electric potentials are measured by electrodes that are either placed between cells in brain tissue (spikes, LFPs), at the cortical surface (ECoG, electrocorticography), or at the scalp (EEG, electroencephalograpy) (Figure~\ref{fig:multimodal}). Spikes are reliable signatures of neuronal action potentials, and spike measurements have been instrumental in mapping out, for example, receptive fields accounting for how sensory stimuli are represented in the brain. The analysis of the LFP signal, the low-frequency part of electric potentials recorded inside gray matter, as well as the ECoG, and EEG signals is less straightforward.
While it is clear that these signals reflect (and therefore contain valuable information about) the underlying neural activity \citep{Einevoll2013,Cohen2017,Pesaran2018}, interpretation of these signals in terms of the underlying neural activity has been difficult. So far, most analyses of LFP, ECoG and EEG data have therefore been purely statistical~\citep{Nunez2006,Buzsaki2012,Einevoll2013,Ilmoniemi2019}. 

There are however good reasons to believe that much could be gained by moving towards a more mechanistic understanding of LFP, ECoG and EEG signals, similar to what has been the tradition in physics. Here candidate hypotheses are typically formulated as specific mathematical models, and predictions computed from the models are compared with experiments. In neuroscience this approach has been used to model activity in individual neurons using, for example, biophysically detailed neuron models based on the cable equation formalism (see, e.g., \citet{Koch1999,Sterratt2011}). These models have largely been developed and tested by comparison with membrane potentials recorded by intracellular electrodes in 
\emph{in vitro} settings (but see \citet{Gold2007}).
To pursue this mechanistic approach to network models in layered structures such as cortex or hippocampus, 
one would like to compare model predictions with all available experimental data, that is, not only spike times recorded for a small subset of the neurons, but also population measures such as LFP, ECoG and EEG signals~\citep{Einevoll2019}. This chapter addresses how to model such electric population signals from neuron and network models.

In addition to allowing for validation on large-scale network models mimicking specific biological 
networks, e.g., \citet{Reimann2013,Markram2015,Billeh2020}, we believe a key application is to generate  
model-based benchmarking data for validation of data analysis methods~\citep{Denker2012}.
One example is the use of such benchmarking data to develop and test spike-sorting methods~\cite{Hagen2016, Buccino2019}
or test methods for localization and classification of cell types~\citep{DelgadoRuz2014, Buccino2018}.
Another example is testing of methods for analysis of LFP signals, such as CSD analysis~\citep{Pettersen2008,Leski2011,Ness2015} or 
ICA analysis~\citep{Glabska2014}, or joint analysis of spike and LFP signals such as laminar population analysis (LPA)~\citep{Glabska2016}.

The standard way to compute extracellular potentials from neural activity is a two-step process~\citep{Holt1999,Linden2014,Hagen2018}:
\begin{enumerate}
\item Compute the net transmembrane current in all neuronal segments in (networks of) biophysically-detailed neuron models, and
\item use volume-conductor (VC) theory to compute extracellular potentials from the these computed transmembrane currents.
\end{enumerate}
In this book chapter we describe the origin of VC theory, that is, how it can be derived from a more detailed electrodiffusive theory describing dynamics of ions in the extracellular media. We further provide examples where our tool LFPy (\texttt{LFPy.github.io}) \citep{Linden2014,Hagen2018} is used to compute spikes, LFP signals and EEG signals generated by neurons and neuronal populations.

\section{
From electrodiffusion to volume conductor theory}
\label{sec:theory}
In this section we describe the origin of extracellular potentials from fundamental electrodiffusive processes. The mathematical derivations we go through can be challenging for people without schooling in mathematics or physics, and readers which are not interested in these mathematical details can consider jumping ahead to section \ref{sec:VC_theory}. We have however attempted to supply some intuitive understanding of what
the different equations represent.

Extracellular potentials are generated by electric currents in the extracellular space. The currents are in turn mediated by movement of ions, and can in principle include several components:
\begin{enumerate}
\item a drift component (ions migrating in electric fields), 
\item a diffusion component (ions diffusing due to concentration gradients),
\item an advective component (extracellular fluid flow drags ions along), and
\item a displacement component (ions pile up and changes the local charge density). 
\end{enumerate}
Since the extracellular bulk fluid has very fast relaxation times and is very close to electroneutral, the latter two current components (3-4) are extremely small and are typically neglected \citep{Grodzinsky2011, Gratiy2017}. The diffusive component (2) is acknowledged to play an important role for voltage dynamics on a tiny spatial scale, such as in synaptic clefts or in the close vicinity of neuronal membranes, where ion concentrations can change dramatically within very short times \citep{Holcman2015, Savtchenko2017, Pods2017}. At the macroscopic tissue level, it is commonly assumed that the diffusive current is much smaller than the drift current, so that in most studies, only the drift component (1) is considered. The extracellular medium can then be treated as a volume conductor (VC), which greatly simplifies the calculation of extracellular potentials~\citep{Holt1999, Linden2014}.

However, if large ion-concentration gradients are present, diffusive currents could in principle affect measurable extracellular potentials \citep{Halnes2016, Halnes2017, Solbra2018}. Thus in scenarios involving dramatic shifts in extracellular concentrations, such as spreading depression and related pathologies, diffusive effects are likely to be of key importance for shaping the extracellular potential \citep{Almeida2004, OConnell2016}. For such cases VC theory is insufficient, and computationally much more expensive electrodiffusive modeling must be used.

\subsection{Ion concentration dynamics}
\label{sec:eldiff}
In this section the starting point is the general assumption of ion movement under the combined influence of electric fields and concentration gradients. Building on this, we first describe computational schemes for modelling electrodiffusive processes, and next show how the electrodiffusive theory reduces to the fundamental equations for VC theory when we assume negligible effects from diffusion.

The movement of ions in the brain are described in terms of fluxes \citep{Freeman1975}. In electrodiffusive processes, the flux density of an ion species $k$ is given by (see e.g., \cite{Grodzinsky2011}):
\begin{equation}
{\bf j_k} = - D_{k} {\bf \nabla} c_{k} - \frac{D_k z_k c_k}{\psi} {\bf \nabla} \phi.
\label{eq:JNP}
\end{equation}
The mathematical operator $\nabla$ computes the spatial derivative (gradient) of scalars.
The first term on the right is Fick's law for the diffusive flux density $j_{k}^\text{diff}$, 
and it implies that the diffusive flux is proportional to the gradient of the concentration, $c_k$.
The second term is the drift flux density $j_{k}^\text{drift}$,
and it implies that the drift flux is proportional to the gradient of the voltage, $\phi$.
This equation expands Fick's law in the case where the diffusing particles also move due to electrostatic forces with a mobility $D_k/\psi$ (cf. the Einstein-relation, \cite{Grodzinsky2011}). Here $D_{k}$ is the diffusion coefficient of ion species $k$, $\phi$ is the electric potential, $z_{k}$ is the valency of ion species $k$, and $\psi=RT/F$ is defined by the gas constant ($R$), Faraday's constant ($F$)  and the temperature ($T$). The ion concentration dynamics of a given species is then given by the Nernst-Planck continuity equation, 
\begin{equation}
\frac{\partial c_k}{\partial t} = - {\bf \nabla} \cdot {\bf j_k} + f_k = {\bf \nabla} \cdot \left[ D_k {\bf \nabla} c_k + \frac{D_k z_k c_k}{\psi} {\bf \nabla} \phi \right] + f_k,
\label{eq:NP}
\end{equation}
where $f_k$ represents any source term in the system, such as e.g., an ionic transmembrane current source \citep{Solbra2018}. 
Note that $\nabla \cdot$ computes the divergence of vectors (such as flux densities), which represents the volume density of the outward flux from an infinitesimal volume around a given point.
Essentially, this equation just tells us that if there is a net movement of ions into or out of a volume (from any source within the volume ($f_k$), and/or moving in from the sides (${\bf \nabla} \cdot {\bf j_k}$)), then the local concentration ($c_k$) must change (left hand side of equation).

In order to solve a set (i.e., one for each ion species present) of equations like eq.~\ref{eq:NP}, one needs an expression for the electric potential $\phi$. There are two main approaches to this. The physically most detailed approach is to use the Poisson-Nernst-Planck (PNP) formalism \citep{Leonetti1998, Leonetti2004, Lu2007, Lopreore2008, Nanninga2008, Pods2013, Gardner2015, Cartailler2018}. Within this formalism, $\phi$ is determined from Poisson's equation from electrostatics, 
\begin{equation}
\nabla^2 \phi = -\rho/\epsilon.
\label{eq:poisson}
\end{equation}
Here $\nabla^2$ is called the Laplacian. The Laplacian of a function is the sum of second partial derivatives with respect to each independent variable. Poisson’s equation is used to find the electric potential arising from a given charge distribution.
Further, $\epsilon$ is the permittivity of the system, and $\rho$ is the charge density associated with the ionic concentrations, as given by
\begin{equation}
\rho = F \sum_k z_k c_k.
\label{eq:F}
\end{equation}
An alternative, more computationally efficient approach is to replace the Poisson equation with the simplifying approximation that the bulk solution is electroneutral \citep{Mori2008, Mori2009, Mori2009a, Mori2011, Niederer2013, Halnes2013, Halnes2015, Pods2017,  OConnell2016, Solbra2018, Tuttle2019, Ellingsrud2020, Saetra2020}, which is a good approximation on spatiotemporal scales larger than micrometers and microseconds \citep{Grodzinsky2011, Pods2017, Solbra2018}. 

Both the PNP formalism and the electroneutral formalism allow us to compute the dynamics of ion concentrations and the electric potential in the extracellular space of neural tissue containing an arbitrary set of neuronal and glial current sources. For example, in recent work, a version of the electroneutral formalism called the Kirchhoff-Nernst-Planck (KNP) formalism was developed into a framework for computing the extracellular dynamics (of $c_k$ and $\phi$) in a 3D space surrounding morphologically complex neurons simulated with the NEURON simulation tool \citep{Solbra2018}. However, both the PNP and electroneutral formalisms such as KNP keep track of the spatial distribution of ion concentrations, and as such they require a suitable meshing of the 3D space, and numerical solutions based on finite difference- or finite element methods. In both cases, simulations can become computationally demanding, and for systems at a tissue level the required computational demand may become unfeasible. For that reason, there is much to gain from deriving a simpler framework where effects of ion concentration dynamics are neglected, and for many scenarios this may be a good approximation. Below, we will derive this simpler framework, i.e., the standard volume conductor (VC) theory, using the Nernst-Planck fluxes (eq.~\ref{eq:JNP}) as a starting point.

\subsection{Electrodynamics}
If we multiply eq.~\ref{eq:JNP} by $F\cdot z_k$ and sum over all ion species, we get an expression for the net electric current density due to all particle fluxes,

\begin{equation}
{\bf i} = - \sum_k{F z_k D_{k}{\bf \nabla} c_{k}} - \sigma {\bf \nabla}{\phi},
\label{eq:INP}
\end{equation}
where the first term is the diffusive current density $i^\text{diff}$ and the second term is the drift current density $i^\text{drift}$. We have here identified the conductivity $\sigma$ of the medium as \citep{Grodzinsky2011}:
\begin{equation}
\sigma = F\sum_{k} \frac{\tilde{D}_{k} z_{k}^2}{\psi}c_{k}.
\label{eq:sigma}
\end{equation}
Current conservation in the extracellular space implies that:
\begin{equation}
{\bf \nabla} \cdot {\bf i} = - \sum_k{F z_k D_{k}\nabla^2 c_{k}} - {\bf \nabla} \cdot (\sigma {\bf \nabla} \phi) = - C,
\label{eq:CSD}
\end{equation}
where $C$ denotes the current source density (CSD). 
This equation implies that because of current conservation, the net amount of current entering or leaving through the sides of a certain volume of the extracellular space (${\bf \nabla} \cdot {\bf i}$), must be exactly balanced by
the net amount of current entering or leaving through current sources and sinks within the volume (-$C$). Here $C$ will reflect e.g., local neuronal or glial transmembrane currents.
We note that this is essentially equivalent to eq.~\ref{eq:NP} at the level of single ion species, with the exception that eq.~\ref{eq:NP} contains a term $\partial c_k/ \partial t$ for accumulation of ion species $k$, while eq.~\ref{eq:CSD} does \emph{not} contain a corresponding term ($\partial \rho/ \partial t$) for charge accumulation. Hence, in eq.~\ref{eq:CSD} it is implicitly assumed that the extracellular bulk solution is electroneutral \citep{Solbra2018}. We note that in general, the $CSD$ term includes both ionic transmembrane currents and transmembrane capacitive currents, and that the latter means that the local charge accumulation building up the transmembrane potential still occurs in the membrane Debye-layer.

Note that if we assume all concentrations to be constant in space, the diffusive term vanishes, and eq.~\ref{eq:CSD} reduces to
\begin{equation}
{\bf \nabla} \cdot (\sigma {\bf \nabla} \phi) = - C.
\label{eq:CSDstandard}
\end{equation}
Importantly, this equation links the measurable extracellular potentials directly to the neuronal transmembrane currents, and it can therefore be used to calculate extracellular
potentials from a given set of neural current sources.
This is also the standard expression used in CSD theory \citep{Mitzdorf1985, Nicholson1975, Pettersen2006}, where spatially distributed recordings of $\phi$ are used to make theoretical predictions of underlying current sources. When using eq.~\ref{eq:CSDstandard}, it is implicitly assumed that the Laplacian of $\phi$ exclusively reflects transmembrane current sources, and that it is not contributed to by diffusive processes. 

Note that there are two commonly used conventions for defining the variables in eqns.~\ref{eq:JNP}-\ref{eq:CSDstandard}. The variables can be defined either relative to a tissue reference volume or relative to an extracellular reference volume. The former convention is the common convention used in volume conductor theory. For this convention, concentrations denote the number of extracellular ions per unit tissue volume, sources denote the number of ions or the net charge per unit tissue volume per second, and flux or current densities are defined per unit tissue cross-section area. Finally, $\sigma$ interprets as the tissue-averaged extracellular conductivity, i.e., it is not the conductivity of the extracellular solution as such, but accounts for the fact that extracellular currents at the coarse-grained scale (i) have tortuous trajectories around neural and glial obstacles, and (ii) are mostly confined to move only through the extracellular fraction (typically about 0.2) of the total tissue volume \citep{Nicholson1998, Nunez2006}.

As eq.~\ref{eq:CSD} indicates, also diffusive processes can in principle contribute to the Laplacian of $\phi$, and if present, they could give rise to a non-zero Laplacian of $\phi$ even in the absence of neuronal sources ($C=0$). Previous computational studies have predicted that effects of diffusion on extracellular potentials are not necessarily small, but tend to be very slow, meaning that they will only affect the very low-frequency components of $\phi$ \citep{Halnes2016, Halnes2017}. This is due to the diffusive current being a direct function of ion concentrations $c_k$, which on a large spatial scale typically vary on a much slower time scale (seconds to minutes) than the fluctuations in $\phi$ that we commonly are interested in (milliseconds to seconds). Furthermore, electrodes used to record $\phi$ typically have a lower cutoff frequency between 0.1 and 1~Hz \citep{Einevoll2013}, which means that most of the tentative diffusive contribution will be filtered out from experimental recordings. It may therefore be a good approximation to neglect the diffusive term, except in the case of pathologically dramatic concentration variations. For the rest of this chapter, we shall do so, and assume that electrodynamics in neural tissue can be determined by eq.~\ref{eq:CSDstandard}.

\subsection{Volume conductor theory}
\label{sec:VC_theory}
In simulations of morphologically complex neurons, one typically computes a set of transmembrane current sources for each neuronal segment \citep{Koch1999}. By assuming that the tissue medium can be approximated as a volume conductor \citep{Holt1999, Linden2014}, one can then use the standard CSD equation (eq.~\ref{eq:CSDstandard}) to perform a forward modeling of the extracellular potential at each point in space surrounding the neuron(s). Since extracellular potentials are generally much smaller than the membrane potential of $\sim$-70~mV, it is common to assume that the neurodynamics is
not affected by extracellular potentials, and to simulate the neurodynamics as a first independent step, before computing the extracellular potentials in the next step. 

If we consider the simple case of a single point-current source $I_1$ at the origin in an isotropic medium, the current density ${\bf i} = -\sigma {\bf \nabla} \phi$ through a spherical shell with area $4\pi r^2$ must, due to the spherical symmetry, equal $I_1/4\pi r^2~{\bf \hat{r}}$. 
Integration with respect to $r$ gives us:
\begin{equation}
\phi = \frac{I_1}{4\pi \sigma r},
\label{eq:pointsource}
\end{equation}
where $r$ is the distance from the source. 

If we have several point-current sources, $I_{1}, I_2, I_3, ... $, in locations ${\bf r}_1, {\bf r}_2, {\bf r}_3 ... $, their contributions add up due to the linearity assumption (see sec.~\ref{sec:VC_assumptions}), and the potential in a point ${\bf r}$ is given by:
\begin{equation}
\phi({\bf r}) = \frac{I_1}{4\pi  \sigma {\bf |r-r}_1|} + \frac{I_2}{4\pi  \sigma {\bf |r-r}_2| } + \frac{I_3}{4\pi  \sigma {\bf |r-r}_3| } + ... = \sum_k \frac{I_k}{4\pi  \sigma {\bf |r-r}_k| }.
\label{eq:VCtheory}
\end{equation}
Eq.~\ref{eq:VCtheory} is often referred to as the point-source approximation \citep{Holt1999, Linden2014}, since the membrane current from a neuronal segment is assumed to enter the extracellular medium in a single point. An often used further development is obtained by integrating eq.~\ref{eq:VCtheory} along the segment axis, corresponding to the transmembrane current being evenly distributed along the segment axis, giving the line-source approximation \citep{Holt1999, Linden2014}.

\subsubsection{Current-dipole approximation}
When estimating the extracellular potential far away from a volume containing a combination of current sinks and sources, 
it can often be useful to express eq.~\eqref{eq:VCtheory} in terms of a multipole expansion. 
That is, $\phi$
can be precisely described by \citep{Nunez2006},
\begin{equation}\label{eq:multipole}
\phi(R) = \frac{C_\text{monopole}}{R} + \frac{C_\text{dipole}}{R^2} + \frac{C_\text{quadrupole}}{R^3} + \frac{C_\text{octupole}}{R^4} + ... ,
\end{equation}
when the distance $R$ from the center of the volume to the measurement point is larger than the distance from volume center to the most peripheral source \citep{Jackson1998}. 

In neural tissue, there will be no current monopole contribution to the extracellular potential,  
$C_\text{monopole}=0$. This follows from the requirement inherent in the cable equation that the sum over all transmembrane currents, including the capacitive currents, across the neuronal surface has to be zero at all points in time~\citep{Pettersen2012}.
Further,
the quadrupole, octupole and higher-order contributions decay rapidly with distance $R$. Consequently, the multipole expansion can be approximated by the dipole contribution for large distances, a simplification known as the current-dipole approximation \citep{Nunez2006}:

\begin{equation}\label{eq:CDA}
\phi(\mathbf{R}) \approx \frac{C_\text{dipole}}{R^2} = \frac{1}{4 \pi \sigma} \frac{|\mathbf{p}| \cos \theta}{R^2}.
\end{equation}
Here, $\mathbf{p}$ is the current dipole moment and $\theta$ is the angle between the current dipole moment and the distance vector $\mathbf{R}$. The current dipole moment can be found by summing up all the position-weighted transmembrane currents from a neuron \citep{Pettersen2008, Pettersen2014, Nunez2006}: 
\begin{equation}\label{eq:dipole}
\mathbf{p} = \sum_{k=1}^N I_k \mathbf{r}_k.
\end{equation}

In the case of a two-compartment neuron model (see Section~\ref{sec:ExtracellularPotentials}) with a current sink $-I$ at location $\mathbf{r}_1$ and a current source $I$ at location $\mathbf{r}_2$, the current dipole moment can be formulated as $\mathbf{p} = -I\mathbf{r}_1 + I\mathbf{r}_2 = I(\mathbf{r}_2 - \mathbf{r}_1) = I\mathbf{d}$, where $\mathbf{d}$ is the distance vector between the current sink and the current source, giving the dipole length $d$ and direction of the current dipole. The current-dipole approximation is applicable in the far-field limit, that is when $R$ is much larger than the dipole length. For an investigation of the applicability of this approximation for the LFP generated by a single neuron, 
see \cite{Linden2010,Naess2021}.

\

\subsubsection{Assumptions in volume conductor theory}
\label{sec:VC_assumptions}
\begin{figure}[!ht]
\begin{center}
\includegraphics[width=0.6\textwidth]{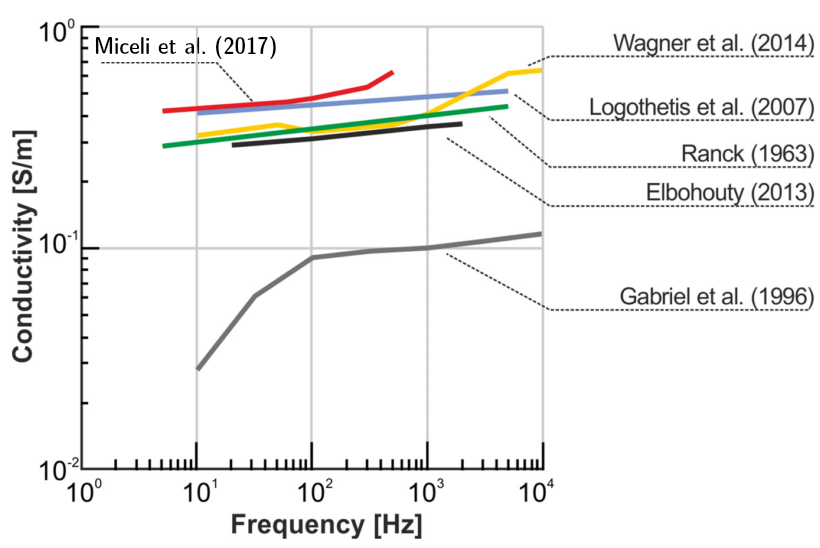}
\end{center}
\caption{\textbf{Literature review of reported conductivities in various species and experimental setups.} 
Most studies seem to indicate a very weak frequency dependence of the extracellular conductivity, which would have a negligible effect on measured extracellular potentials \citep{Miceli2017}. The very low and strongly frequency dependent values measured by \cite{Gabriel1996} represents an outlier, and although it has received substantial attention, it has to the best of our knowledge not been reproduced by any other study.
For details about the data, see \citep{Miceli2017}, and references therein \citep{Ranck1963, Gabriel1996, Logothetis2007, Elbohouty2013, Wagner2014}.
}
\label{fig:freq_dep}
\end{figure}

The point-source approximation, eq.~\ref{eq:VCtheory} (or the line-source version of it), and the current-dipole approximation, eq.~\eqref{eq:CDA}, represent volume conductor theory in its simplest form, and are based on a set of assumptions, some of which may be relaxed for problems where it is relevant: 

\begin{enumerate}

\item {\bf Quasi-static approximation of Maxwell's equations:} Terms with time derivatives of the electric and magnetic fields are neglected. This approximation appears to be well-justified for the relatively low frequencies relevant for brain signals, below about 10 kHz \citep{Nunez2006}.

\item {\bf Linear extracellular medium:} Linear relationship (${\bf i} = -\sigma {\nabla \phi}$) between the current density ${\bf i}$ and the electric field, $\nabla \phi$. This is essentially Ohm's law for volume conductors, and the relation is constitutive, meaning that it is observed in nature rather than derived from any physical principle \citep{Nunez2006, Pettersen2012}.

\item {\bf Frequency-independent conductivity:} Capacitive effects in neural tissue are assumed to be negligible compared to resistive effects in volume conduction. This approximation seems to be justified for the relevant frequencies in extracellular recordings \citep{Logothetis2007, Miceli2017, Ranta2017}, see Fig.~\ref{fig:freq_dep}. Note that it is possible to expand the formalism to include a frequency-dependent conductivity \citep{Tracey2011, Miceli2017}. 

\item {\bf Isotropic conductivity:} The electric conductivity, $\sigma$, is assumed to be the same in all spatial directions. 
Cortical measurements have indeed found the conductivities to be comparable across different lateral directions in cortical grey matter \citep{Logothetis2007}. However, the conductivity in the depth direction, i.e., parallel to the long apical dendrites, was found to be up to 50\% larger than in the lateral direction in rat barrel cortex \citep{goto2010}. Anisotropic electric conductivities have also been found in other brain regions, for example in frog cerebellum \citep{Nicholson1975} and in guinea-pig hippocampus \citep{holsheimer1987}. The approximation that $\sigma$ is homogeneous is still often acceptable, as it normally gives fairly good estimates of the extracellular potential, at least in cortical tissue \citep{Ness2015}. However, it is relatively straightforward to expand the formalism to account for anisotropic conductivities \citep{Ness2015}.


\item {\bf Homogeneous  conductivity:} The extracellular medium was assumed to have the same conductivity everywhere. Although neural tissue is highly non-homogeneous on the micrometer scale \citep{Nicholson1998}, microscale inhomogeneities may average out on a larger spatial scale, and a homogeneous conductivity seems to be a reasonable approximation within cortex \citep{Logothetis2007}. In hippocampus, however, the conductivity has been found to be layer-specific \citep{lopez2001}. In situations where the assumption of a homogeneous conductivity is not applicable, eq.~\ref{eq:CSDstandard} can always be solved for arbitrarily complex geometries using numerical methods, like the Finite Element Method (FEM) \citep{Logg2012}. For some example neuroscience applications, see \cite{Moffitt2005, Frey2009, Joucla2012, Haufe2015, Ness2015, Buccino2019b, Obien2019}. For some simple non-homogeneous cases analytical solutions can still be obtained, for example through the Method of Images for {\it in vitro} brain slices \citep{Ness2015}, and the four-sphere head model for EEG signals (Sec.~\ref{sec:EEG}) \citep{Naess2017}.

\item {\bf No effects from ion diffusion:} To account for diffusion of ions, one would need to compute the electrodynamics of the system using one of the electrodiffusive frameworks presented in Section \ref{sec:eldiff}.

\end{enumerate}

Volume conductor theory is the fundament for forward modeling of extracellular potentials at different spatial scales, from extracellular spikes, LFPs and MUAs, to ECoGs and EEGs. In the following sections we shall review previous modeling works, and insights from simulating electric potentials at these different scales.
We use the software LFPy \citep{Linden2014, Hagen2018, Hagen2019}, which has volume conductor theory incorporated and can in principle be used to compute extracellular potentials on arbitrarily large spatial scales, surrounding arbitrarily large neuronal populations. 

\subsection{Modeling electrodes}

The simplest and most commonly used approach when modeling extracellular recordings is to calculate the extracellular potential at single points following one of the approaches outlined above, and use this as a measure of recorded potentials. Implicitly, this 
assumes ideal point electrodes, that is, the electrodes (and electrode shank) do not affect the extracellular potential and the extracellular potential does not vary substantially over the surface of the electrodes. (The point-electrode assumption was used for all simulation examples in this chapter).

A numerically straightforward extension is the disc-electrode approximation where the potential is evaluated at a number of points on the electrode surface, and the average calculated \citep{ Linden2014}. 
This approach takes into account the physical extent of the electrode, but not any effect the electrode itself might have on the electric potential. 
Close to the electrode surface the electric potential will however be affected by the presence of the high-conductivity electrode contact \citep{McIntyre2001, Moulin2008}. A numerically much more comprehensive approach to modeling electrodes is to use the Finite Element Method (FEM) to model the electrode \citep{Moulin2008, Ness2015}, or the electrode shank \citep{Moffitt2005, Buccino2019b}. Using FEM for validation, \cite{Ness2015} found that the ideal point-electrode and disc-electrode approximations where reasonably accurate when the distance between the current sources and the recording electrode was bigger than $\sim$4 times and $\sim$2 times the electrode radius, respectively, indicating that the effects of the electrodes themselves are negligible in most cases \citep{Nelson2010}.
The presence of large multi-contact electrode probes can, however, substantially affect the extracellular potential in its vicinity, by effectively introducing a large non-conducting volume~\citep{Mechler2012}, and this can amplify or dampen recorded potentials from nearby cells by almost a factor of two, depending on whether the cell is in front of or behind the electrode shank \citep{Buccino2019b}.

Note that for modelling current stimulation electrodes (as opposed to just recording electrodes), more complex electrode models might be needed due to electrode polarization effects \citep{McIntyre2001, Martinsen2008, Joucla2012}.

\section{Single-cell contributions to extracellular potentials} \label{sec:ExtracellularPotentials}



\begin{figure}[!ht]
\begin{center}
\includegraphics[width=1\textwidth]{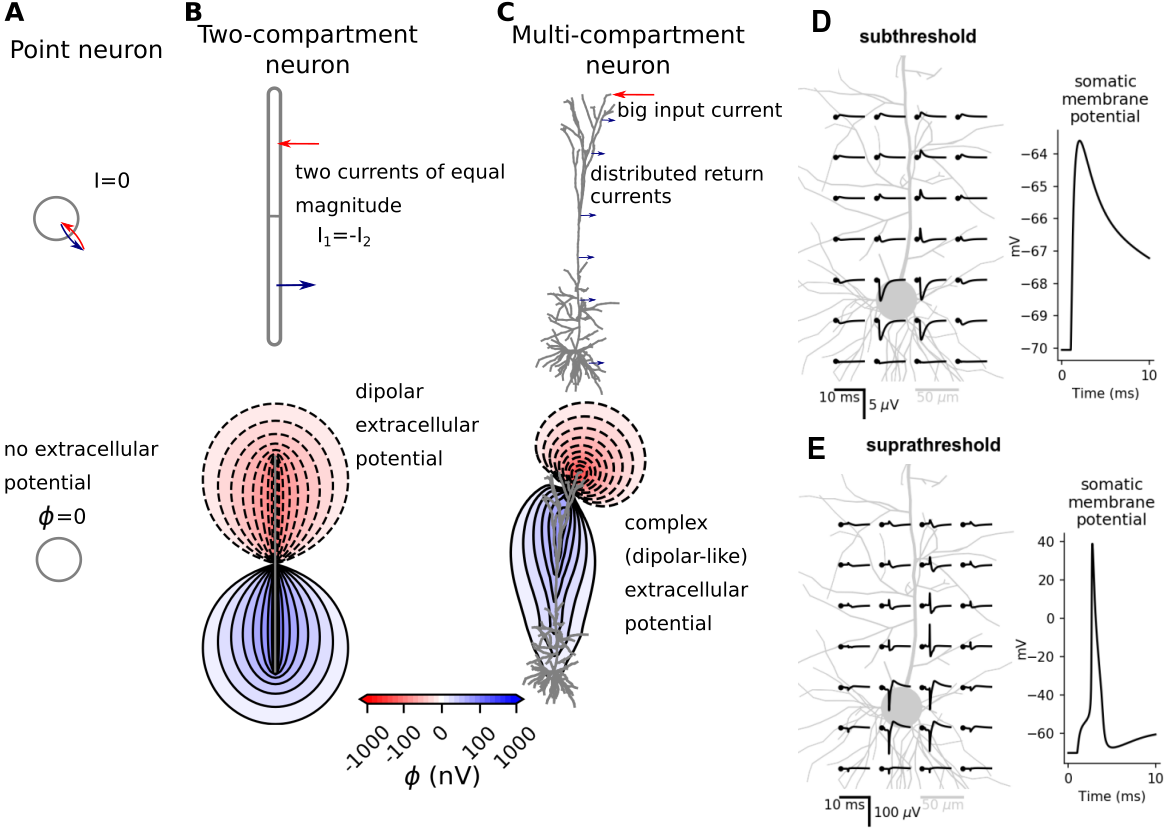}
\end{center}
\caption{\textbf{Single-cell contributions to the extracellular potential.} 
{\bf A}: Point neurons have no net currents (top), and therefore cause no extracellular potentials (bottom). 
{\bf B}: Two-compartment neuron models have two opposite currents
of identical magnitude (top), and cause perfectly symmetric dipolar-like extracellular potentials (bottom). 
{\bf C}: Multi-compartment neuron models \citep{Hay2011} give rise to complex source-sink patterns (top) and complex (but mostly dipolar-like) extracellular potentials (bottom). 
{\bf D, E}: A single somatic synaptic input to a complex multi-compartment cell model, either subthreshold (D) or suprathreshold (E; double synaptic weight of D), illustrating that the same framework can be used to calculate both the extracellular potential from subthreshold synaptic input, and extracellular action potentials. 
}
\label{fig:EP_morph}
\end{figure}

The transmembrane currents of a neuron during any neural activity can be used to calculate extracellular potentials, by applying the formalism described in Sec.~\ref{sec:VC_theory}, and in the simplest case eq.~\ref{eq:VCtheory}.
Current conservation requires that the transmembrane currents across the entire cellular membrane at any given time sum to zero \citep{Koch1999, Nunez2006}, and since
an excitatory synaptic input generates a current sink (negative current), this will necessarily lead to current sources elsewhere on the cell. This implies that point neurons, that is, neurons with no spatial structure, will have no net transmembrane currents, and hence cause no extracellular potentials (Fig.~\ref{fig:EP_morph}A). The simplest neuron models that are capable of producing extracellular potentials are therefore two-compartment models, which will have two equal but opposite transmembrane currents,  
giving rise to perfectly symmetric extracellular potentials (Fig.~\ref{fig:EP_morph}B).


Multi-compartment neuron models mimicking the complex spatial structure of real neurons will typically give rise to complicated patterns of current sinks and sources, leading to complex, but mostly dipolar-like extracellular potentials (Fig.~\ref{fig:EP_morph}C) \citep{Einevoll2013}.
Note that this framework for calculating extracellular potentials is valid both for subthreshold and suprathreshold neural activity, that is, when a cell receives synaptic input that does not trigger, or does trigger an action potential, respectively (Fig.~\ref{fig:EP_morph}, D versus E).

\section{Intra-cortical extracellular potentials from neural populations}
\begin{figure}[!ht]
\begin{center}
\includegraphics[width=1\textwidth]{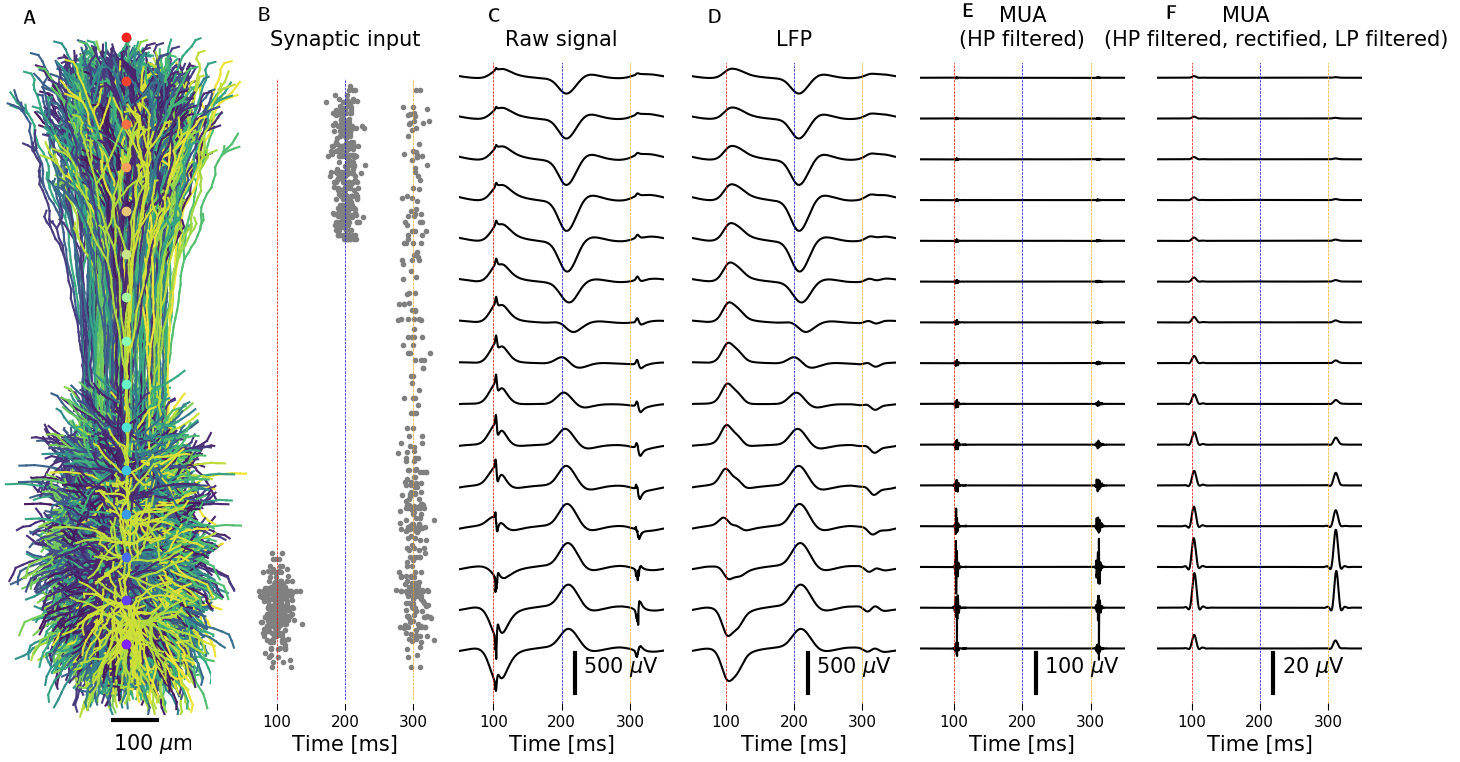}
\end{center}
\caption{\textbf{Extracellular potentials from different waves of synaptic input}. Different brain signals from separate waves of excitatory synaptic input to 10 000 layer 5 pyramidal cells from rat \citep{Hay2011}.
{\bf A}: A subset of 100 pyramidal cells, with the LFP electrode locations indicated in the center (colored dots).
{\bf B}: Depth-resolved synaptic inputs arrive in three waves, first targeting the basal dendrites (t=100~ms), then the apical dendrites (t=200~ms), and lastly uniformly across the entire depth (t=300~ms). Note that all synaptic input is pre-defined, that is, there is no network activity.
{\bf C}: The extracellular potential at different depths (corresponding to dots in panel A), including both spikes and synaptic input.
{\bf D}: The LFP, that is, a low-pass filtered version of the raw signal in C. 
{\bf E}: The MUA, that is, a high-pass filtered version of the raw signal in C.
{\bf F}: Another version of the MUA which is a rectified and low-pass filtered version of the MUA signal in E.
All filters were 4th order Butterworth filters in forward-backward mode \citep{NeuroEnsamble2017}. For illustrative purposes a relatively low cut-off frequency of 50~Hz was chosen for the LFP low-pass filter. The MUA was first high-pass filtered above 300~Hz (E and F), then rectified and low-pass filtered below 300 Hz (F).
}
\label{fig:LFP_MUA}
\end{figure}

Extracellular potentials measured within neural tissue are often split into two separate frequency domains, which reflect different aspects of the underlying neural activity. 
The low frequency part, the local field potential (LFP), is thought to mostly reflect synaptic input to populations of pyramidal cells, while the high-frequency part, the multi-unit activity (MUA), reflects the population spiking activity (Fig.~\ref{fig:LFP_MUA}).

\subsection{Local field potentials}
The LFP is the low-frequency part ($\lesssim$ 500~Hz) of the extracellular potentials, and it is among the oldest and most used brain signals in neuroscience \citep{Einevoll2013}. The LFP is expected to be dominated by synaptic inputs asymmetrically placed onto populations of geometrically aligned neurons \citep{Nunez2006, Linden2011, Einevoll2013b}.
In cortex and hippocampus, neurons can broadly speaking be divided into two main classes: the inhibitory interneurons, and the excitatory pyramidal neurons. Pyramidal neurons typically have a clear axis of orientation, that is, the apical dendrites of close-by pyramidal neurons tend to be oriented in the same direction (Fig.~\ref{fig:LFP_MUA}A). This geometrical alignment is important because the LFP contributions from the individual pyramidal cells also align and therefore sum constructively.  For example, basal excitatory synaptic input (Fig.~\ref{fig:LFP_MUA}B, time marked by red line) generates a current sink and corresponding negative LFP deflection in the basal region, and simultaneously a current source and corresponding positive LFP deflection in the apical region (Fig.~\ref{fig:LFP_MUA}D, time marked by red line), while apical excitatory synaptic input leads to the reversed pattern (Fig.~\ref{fig:LFP_MUA}B,D, time marked by blue line). 
Importantly, this means that excitatory input that simultaneously targets both the apical and the basal dendrite will give opposite source/sink patterns which will lead to substantial cancellation and a weak LFP contribution (Fig.~\ref{fig:LFP_MUA}B, D, time marked by orange line).
The same arguments also apply to inhibitory synaptic inputs, with the signs of the currents and LFPs reversed. 

Note that, for example, the LFP signature of apical excitatory synaptic input is inherently similar to that of basal inhibitory input, and indeed, separating between cases like this pose a real challenge in interpreting LFP signals \citep{Linden2010}. 

In contrast to pyramidal neurons, interneurons often lack any clear orientational specificity, meaning that the current dipoles from individual interneurons, which might by themselves be sizable \citep{Linden2010}, do not align, leading to negligible net contributions to LFP 
signals~\citep{Mazzoni2015,Martinez-Canada2021}.
Note, however, that the interneurons may indirectly cause large LFP contributions through their 
synaptic inputs onto pyramidal cells \citep{Telenczuk2016, Hagen2016}.

It has been demonstrated that correlations among the synaptic inputs to pyramidal cells can amplify the LFP signal power by orders of magnitude, with the implication that populations receiving correlated synaptic input can dominate the LFP also 1-2~mm outside of the population \citep{Linden2011, Leski2013}.

Somatic action potentials lasting only a few milliseconds are generally expected to contribute little to cortical LFP signals 
\citep{Pettersen2008,Pettersen2008a,Einevoll2013, Haider2016}:  Their very short duration with both positive and negative phases (Fig.~\ref{fig:EP_morph}E) will typically give large signal cancellations of the contributions from individual neurons, and their high frequency content is to a large degree removed from LFPs during low-pass filtering. Note, however, that in the hippocampus the highly synchronized spikes found during sharp wave ripples are expected to also contribute to shaping of the LFP \citep{Schomburg2012, Luo2018}.

Other active conductances may contribute in shaping the LFP, for example, the slower dendritic calcium 
spikes \citep{Suzuki2017} or long-lasting after-hyperpolarization currents~\citep{Reimann2013}. Further, 
subthreshold active conductances can also shape the LFP by molding the transmembrane currents following synaptic input, and the hyperpolarization-activated cation channel I$_{\rm h}$ may play a key role in this, both through asymmetrically changing the membrane conductance, and by introducing apparent resonance peaks in the LFP \citep{Ness2016, Ness2018}.

\subsection{MUA}
While LFPs are thought to mainly reflect the synaptic input to large populations of pyramidal neurons, the multi-unit activity (MUA) can be used to probe the population spiking activity \citep{Einevoll2007,Pettersen2008} (Fig.~\ref{fig:LFP_MUA} E,F). In other words, the MUA holds complimentary information to the LFP. In particular, this can be useful for some cell-types, like excitatory stellate cells and inhibitory interneurons, which are expected to have very weak LFP contributions \citep{Linden2011}, but might still be measurable through their spiking activity. 
Similarly, spatially uniformly distributed synaptic input to pyramidal neurons results in a negligible LFP contribution (Fig.~\ref{fig:LFP_MUA}C, time marked by orange line), while the population might still contribute substantially to the MUA through the extracellular action potentials (Fig.~\ref{fig:LFP_MUA}E-F, time marked by orange line). 

\section{ECoG and EEG}
\label{sec:EEG}
In order to measure electric potentials in the immediate vicinity of neurons, like LFP and MUA signals, we need to insert electrodes into the brain. This highly invasive technique is quite common in animal studies, but can only be applied to humans when there is a clear medical need, for example in patients with intractable epilepsy \citep{Zangiabadi2019}. However, electric potentials generated by neural activity extend beyond neural tissue and can also be measured outside the brain:
Placing electrodes on the brain surface, as in electrocorticography (ECoG), is a technique that requires surgery. With electroencephalography (EEG), on the other hand, potentials are measured non-invasively, directly on top of the scalp.

\begin{figure}[!ht]
\begin{center}
\includegraphics[width=0.5\textwidth]{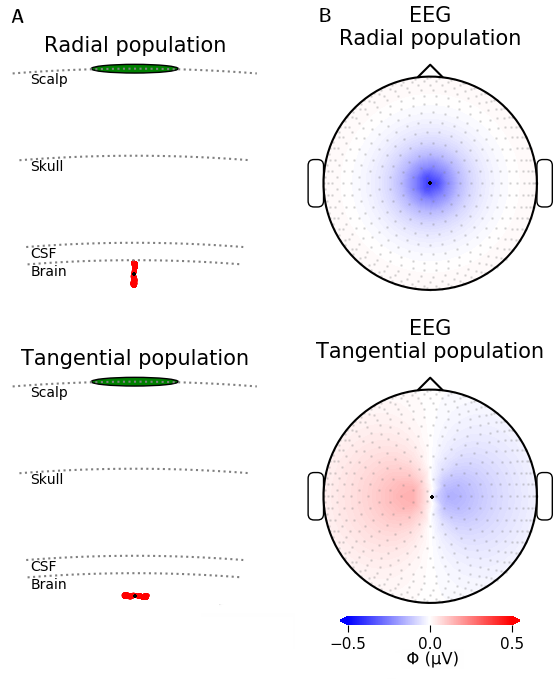}
\end{center}
\caption{\textbf{EEG from apical synaptic input to population of pyramidal cells}.
{\bf A}: The four-sphere head model with two orientations of the neural population from Fig.~\ref{fig:LFP_MUA}, either radial, mimicking a population in a gyrus (top) or tangential, mimicking a population in a sulcus (bottom).
{\bf B}: A snapshot of the EEG signal at the head surface for apical input (time marked with blue dotted line in Fig.~\ref{fig:LFP_MUA}), for a radial population (top) or tangential population (bottom).
The center of the population is marked with a black dot.
}
\label{fig:EEG_MEG}
\end{figure}

Since EEG electrodes are located relatively far away from the neuronal sources, the current dipole approximation, eq.~\eqref{eq:CDA}, combined with some head model, can be applied for computing EEG signals \citep{Nunez2006,Ilmoniemi2019,Naess2021}. By collapsing the transmembrane currents of a neuron simulation into one single current dipole moment, see eq.~\eqref{eq:dipole}, we can calculate EEG from arbitrary neural activity (Fig.~\ref{fig:EEG_MEG}).
The current dipole approximation is however not unproblematic to use for computing ECoG signals, as the ECoG electrodes may be located too close to the 
signal sources for the approximation to apply, see \cite{Naess2021}.

\subsection{Head models}
\begin{figure}[!ht]
\begin{center}
\includegraphics[width=1.0\textwidth]{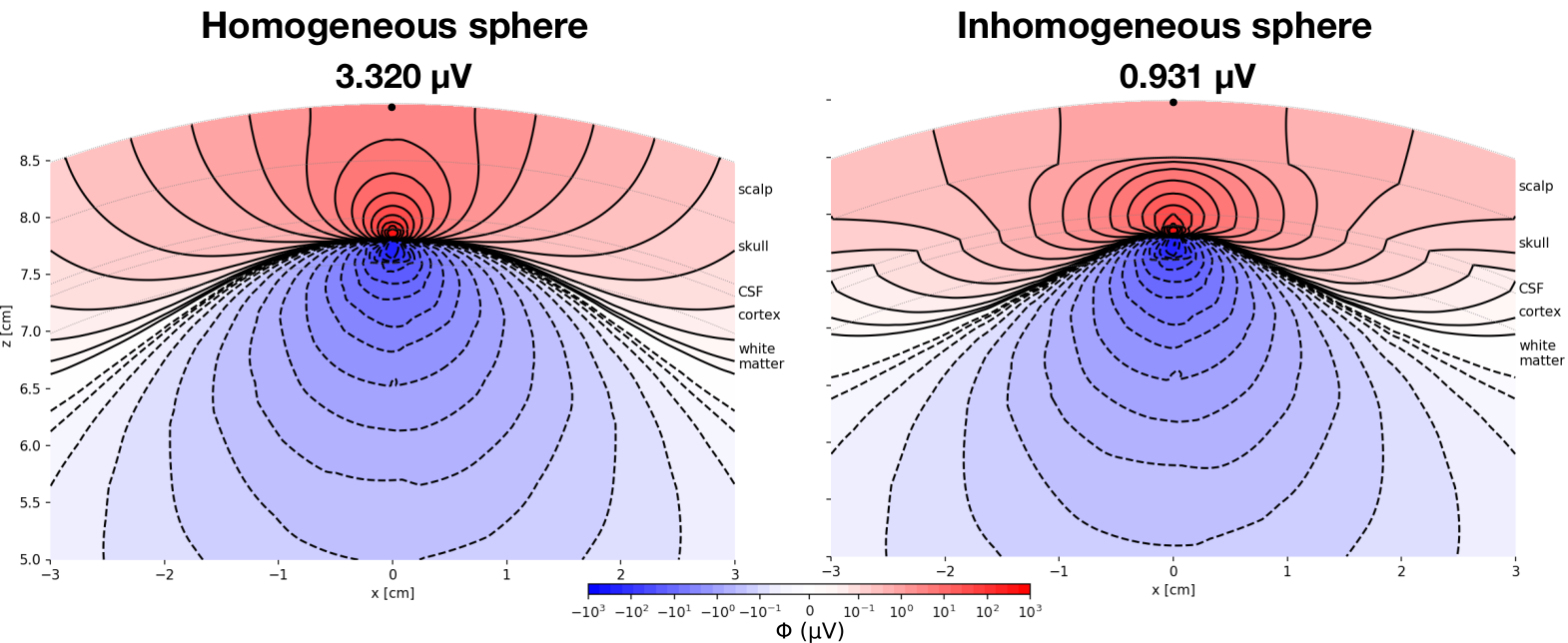}
\end{center}
\caption{\textbf{Effect of head inhomogeneities}.
The same current dipole will give substantially different potentials on the head surface if the different conductivities of the head is included in a FEM model \citep{Naess2017}. Left: Homogeneous sphere, with electrical conductivity, $\sigma=0.33$~S/m everywhere. Right: Standard four-sphere head model, with $\sigma_{\text{brain}}=0.33$~S/m, $\sigma_{\text{CSF}}=5\sigma_{\text{brain}}$, 
$\sigma_{\text{skull}}=\sigma_{\text{brain}} / 20$, $\sigma_{\text{scalp}}=\sigma_{\text{brain}}$.
}
\label{fig:foursphere_contour}
\end{figure}

\begin{figure}[!ht]
\begin{center}
\includegraphics[width=0.6\textwidth]{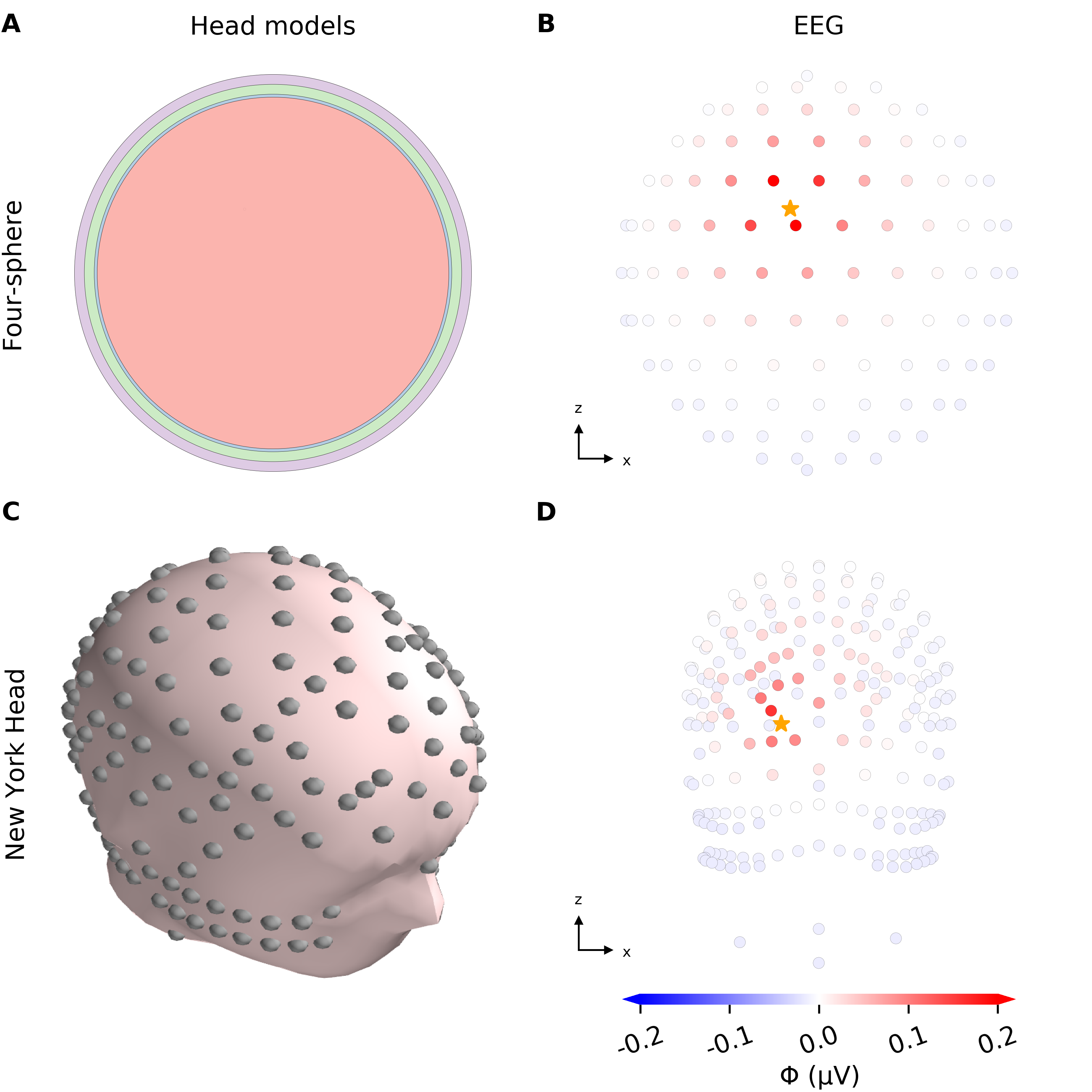}
\end{center}
\caption{\textbf{The four-sphere head model and the NY Head model}. EEG signals from population dipole resulting from waves of excitatory synaptic input to 10 000 layer 5 pyramidal cells from rat \citep{Hay2011}.
	{\bf A}: The four-sphere model consisting of four concentrical shells: brain, CSF, skull and scalp. 
	{\bf B}: Maximum EEG signals ($V_e$) on scalp surface electrodes resulting from population dipole placed at location marked by orange star, computed with the four-sphere model.
	{\bf C}: Illustration of the New York Head model \citep{Huang2016,Naess2021}.
	{\bf D}: EEG signals computed with the New York Head model, equivalent to panel {\bf B}.
}
\label{fig:head_models}
\end{figure}

Electric potentials measured on the scalp surface will be affected by the geometries and conductivities of the different constituents of the head (Fig.~\ref{fig:foursphere_contour}) \citep{Nunez2006}. This can be incorporated in EEG calculations by applying simplified or more complex head models.
A well-known simplified head model is the analytical four-sphere model, consisting of four concentric shells representing brain tissue, cerebrospinal fluid (CSF), skull and scalp, where the conductivity can be set individually for each shell \citep{Naess2017, Srinivasan1998, Nunez2006} (Fig.~\ref{fig:foursphere_contour}, Fig.~\ref{fig:head_models}A,B).
More complex head models make use of high-resolution anatomical MRI-data to map out a geometrically detailed head volume conductor. The link between current dipoles in the brain and resulting EEG signals is determined applying numerical methods such as the finite element method \citep{Larson2013, Logg2012}. Once this link is established we can in principle insert a dipole representing arbitrary neural activity into such a model, and compute the resulting EEG signals quite straightforwardly \citep{Naess2021,Martinez-Canada2021}. The New York Head model is an example of one such pre-solved complex head model, see Fig.~\ref{fig:head_models}C,D \citep{Huang2016}.

The head models themselves introduce no essential frequency filtering of the EEG signal \citep{Pfurtscheller1975, Nunez2006, Ranta2017}, however, substantial spatial filtering will occur (Fig.~\ref{fig:foursphere_contour}). Additionally, the measured (or modeled) signals represent the average potential across the elecrode surface, and the large electrode sizes used in ECoG/EEG recordings can have important effect on the measured signals \citep{Nunez2006, Hagen2018, Dubey2019}.
%
%

%


\section{Conclusion}
\label{sec:summary}

In the present chapter we have derived and applied well-established biophysical forward-modeling schemes for computing extracellular electric potentials recorded inside and outside the brain. These electric potentials include spikes (both single-unit and multiunit activity (MUA)), LFP, ECoG and EEG signals. The obvious application of this scheme is computation of electric signals from neuron and network activity for comparison with experiments so that candidate models can be tested~\citep{Einevoll2019,Martinez-Canada2021} or inferred~\citep{Goncalves2019,Skaar2020}.
Another key application is the computation of benchmarking data for testing of data analysis methods such as 
spike sorting or CSD analysis~\citep{Denker2012}. 

Inverse modeling of recorded electric potentials, that is, estimation of the neural sources underlying the signals, is inherently an ill-posed problem. This means that no unique solution for the size and position of the sources exists. However, prior knowledge about the underlying sources and how they generate the recorded signals, can be used to increase the identifiability. For example, several methods for the estimation of so-called current-source density (CSD) from LFP recordings have been developed by building the present forward model into the CSD 
estimator~\citep{Pettersen2006,Potworowski2012,Cserpan2017}.

The present chapter has focused on the modeling of measurements of extracellular electric signals. There are several other measurement modalities where detailed forward modeling could be pursued to allow for a more quantitative analysis of recorded data, such as magnetoencephalograpy (MEG), where magnetic fields are recorded outside the head, Voltage-sensitive dye imaging (VSDI), which reflects area-weighted neuronal membrane potentials ~\citep{Chemla2012}, two-photon calcium imaging, which measures the intracellular calcium dynamics ~\citep{Helmchen2012} and functional magnetic resonance imaging (fMRI), which reflects blood dynamics ~\citep{Bartels2012}. While blood dynamics is typically not explicitly included in neural network models, MEG, VSDI and calcium imaging are accessible through neuronal simulations of the type used to compute electric signals. Similar to EEG, the MEG stems from the transmembrane currents of neurons and can be computed based on the current dipoles of the underlying neurons ~\citep{Hamalainen1993,Ilmoniemi2019,Naess2021}.The new version of our tool LFPy, which was used in generating the examples in the present chapter, thus also includes the ability to compute MEG signals ~\citep{Hagen2018}.
 



\section*{Acknowledgements}
\label{sec:acknowledgements}
This research has received funding from the European Union Horizon 2020 Framework Programme for Research
and Innovation under Specific Grant Agreement No. 785907 and No. 945539 [Human Brain Project (HBP) SGA2 and SGA3], and the Research Council of Norway (Notur, nn4661k; DigiBrain, no. 248828; INCF National Node, no. 269774).

\label{sec:bibliography}
\bibliography{ECS_bookchapter}


\end{document}